\documentclass[a4paper,pdftex]{amsart}
\usepackage{amsmath}
\usepackage{amssymb}
\usepackage{url}

\newtheorem{theorem}{Theorem}

\newtheorem{lemma}{Lemma}
\newtheorem{proposition}{Proposition}

\newtheorem{definition}{Definition}
\newtheorem{remark}{Remark}

\newtheorem{example}{Example}

\newcommand{\F}{\mathbb{F}}
\newcommand{\N}{\mathbb{N}}

\newcommand{\kernel}{\mbox{\rm Ker \,}}
\newcommand{\dist}{\mbox{\rm d}}
\newcommand{\MRD}{\mbox{\rm MRD}}
\newcommand{\GL}{\mbox{\rm GL}}
\DeclareMathOperator{\rank}{rank}
\DeclareMathOperator{\tr}{tr}
\newcommand{\kker}{\mbox{\rm Ker\,}}
\newcommand{\SL}{\mbox{\rm SL}}

\begin{document}

\title{Algebraic structures of MRD Codes}
\thanks{The authors wish to thank Colciencias, Bogot{\'a}, Universidad del Norte, Barranquilla, and \\ \mbox{} \quad \! COST Action IC 1104 for financial support.}

\author[J. de la Cruz, M. Kiermaier, A. Wassermann, W. Willems]{Javier de la Cruz \and Michael Kiermaier \and Alfred Wassermann \and Wolfgang Willems}

\address{J. de la Cruz
    Departamento de Matem\'aticas, Universidad del Norte, Km 5 V\'ia Puerto Colombia, Barranquilla, Colombia
}
\email{jdelacruz@uninorte.edu.co}

\address{M. Kiermaier 
              Mathematisches Institut, Universit\"{a}t Bayreuth, 95447 Bayreuth, Germany 
	      }
	\email{michael.kiermaier@uni-bayreuth.de}

\address{A. Wassermann 
              Mathematisches Institut, Universit\"{a}t Bayreuth, 95447 Bayreuth, Germany  
}
\email{alfred.wassermann@uni-bayreuth.de}

\address{W. Willems
            Departamento de Matem\'aticas, Universidad del Norte, Km 5 V\'ia Puerto Colombia, Barranquilla, Colombia and
            Institut f\"ur Algebra und Geometrie, Otto-von-Guericke Universit\"at, Postfach 4120, 39016 Magdeburg, Germany
}
\email{willems@ovgu.de}

\maketitle

 \centerline{In memory of Axel Kohnert}
 \medskip

\begin{abstract} Based on results in finite geometry we prove the existence of MRD codes in $(\F_q)_{n,n}$ with minimum distance $n$ which are essentially different from
Gabidulin codes. The construction results from algebraic structures which are closely related to those of finite fields. Some of the results may be known to experts, but to our knowledge have never been pointed out explicitly in the literature.
\end{abstract}

\section{Introduction} Let $\F_q$ denote a finite field with $q$ elements and let $V=(\F_q)_{m,n}$ be the $\F_q$-vector space of 
matrices over $\F_q$ of type $(m,n)$. On $V$ we define the so-called rank metric distance by 
$$ \dist(A,B) =\rank(A-B) $$
for $A,B \in V$. Clearly, the distance $\dist$ is a translation invariant metric on $V$.
A subset $\mathcal{C} \subseteq V$ endowed with the metric $\dist$ is called a rank metric code with minimum distance $$\dist(\mathcal{C})= \min\,\{ \dist(A,B)\mid A \not= B \in V \}.$$
For $m \geq n$, an $\MRD$ (maximum rank distance) code $\mathcal{C} \subseteq V$ satisfies the following two conditions:
\begin{itemize}
\item[\rm (i)] $|\mathcal{C}|=q^{km}$ and
\item[\rm (ii)] $\dist(\mathcal{C}) = n-k+1$.
\end{itemize}
Note that an $\MRD$ code is a rank metric code which is maximal in size given the minimum distance, or in other words it achieves the Singleton bound 
for the rank metric distance (see \cite{D,G}).\\[1ex]
Delsarte was the first who proved in \cite{D}
 the existence of linear $\MRD$ codes for all $q,m,n$ and $ 1 \leq k \leq n$.
His construction (in the notation of Gabidulin \cite{G}) runs as follows:
Let $a_1, \ldots, a_n \in \F_{q^m}$ be linearly independent over $\F_q$ and let $C$ be the $\F_{q^m}$-linear code
defined by the generator matrix
$$ 
G = \left(
\begin{array}{lll}
  a_1 & \ldots & a_n \\
  a_1^q & \ldots & a_n^q \\
  \vdots & \ldots & \vdots \\
  a_1^{q^{k-1}} & \ldots & a_n^{q^{k-1}}
\end{array}
\right), 
$$
where $ 1 \leq k \leq n$.	
Each code word $c \in C$ is a vector in $(\F_{q^m})^n$. If we choose a fixed basis of $\F_{q^m}$ over $\F_q$
then $c$ may be regarded as a matrix in $V=(\F_q)_{m,n}$. Thus we obtain an $\F_q$-linear code $\mathcal{C}$ in $V$. 
The code $\mathcal{C}$, which is usually called a Gabidulin code (although first discovered by Delsarte), is an $\F_q$-linear $\MRD$ code of size $q^{km}$ with minimum distance $d=n-k+1$. 
At this point we may naturally ask: Is there any other MRD code which is essentially different from a Gabidulin code, i.e., which does not allow an isometry to a
Gabidulin code. \\

\begin{definition} {\rm a) A bijective map $\varphi: (\F_q)_{m,n} \longrightarrow (\F_q)_{m,n}$ is called an isometry if $\varphi$ preserves the rank metric distance, i.e., 
$$ \dist(A,B) = \dist (\varphi(A),\varphi(B))$$
for all $A,B \in (\F_q)_{m,n}$. \\
b) Two codes $\mathcal{C}$ and $\mathcal{C}'$ in $(\F_q)_{m,n}$ are equivalent if there exists an isometry $\varphi$ with $\varphi(\mathcal{C}) = \mathcal{C}'$.
If one of the codes is additively closed resp. an $\F_q$-vector space, we require in addition that $\varphi$ is additive resp. $\F_q$-linear.
}
\end{definition}
In odd characteristic, already in the 1950s of the last century L.-K. Hua has classified all bijective maps $\varphi$ from $(\F_q)_{m,n}$ onto itself such that $\varphi$ and $\varphi^{-1}$ preserve the distance between adjacent matrices, i.e., between all pairs of matrices $A,B$ with $\rank (A-B) =1$ \cite{Hua}.
In even characteristic, this has been done by Z.-X. Wan in the 1960s \cite{Wa}.
For isometries the result can be stated as follows (see Theorem 3.4 in \cite{Wan}): 

\begin{theorem}{\rm (Hua, Wan)} \label{Hua} If $\varphi$ is an isometry of $(\F_q)_{m,n}$ with $m,n \geq 2$, then there exist matrices $X \in \GL(m,q), Y \in \GL(n,q)$ and $Z \in (F_q)_{m,n}$ such that
$$ \varphi(A) = XA^\sigma Y + Z \quad \mbox{for all} \ A \in (\F_q)_{m,n} $$
where $\sigma$ is an automorphism of the field $\F_q$ acting on the entries of $A$, \\
or, but only in case $m=n$,
$$ \varphi(A) = X(A^t)^\sigma Y + Z \quad \mbox{for all} \ A \in (\F_q)_{n,n} $$
where $A^t$ denotes the transpose of $A$. \\
If $\varphi$ is additive, then obviously $Z=0$. In addition $\sigma=\operatorname{id}$ in case $\varphi$ is $\F_q$-linear.
\end{theorem}

In the recent paper \cite{M} Morrison has rediscovered Hua's result in case that $\varphi$ is linear resp. semi-linear. 
Applications of $\MRD$-Codes to random matrices over finite fields are discussed in \cite{YH}.

If we specialize the Gabidulin construction to $k=1$ and $m=n$, then a Gabidulin code $\mathcal{G}$ is a linear code of dimension $n$ such that $A-B$ is always regular for $A \not= B$ in $\mathcal{G}$. In particular, all $ 0 \not= A \in \mathcal{G}$ are regular since the zero matrix is in $\mathcal{G}$.
 If $\mathcal{G}$ is defined by $(a_1,\ldots,a_n) \in \F_{q^n}^n$ and we choose $a_1, \ldots, a_n$ as an $\F_q$-basis $B$ of $\F_{q^n}$ then 
$$\mathcal{G} = \langle S \rangle \cup \{0\}$$
 where $\langle S \rangle$ denotes the group generated by a Singer cycle $S$ in $\GL(n,q)$
which is the matrix defining the multiplication by a primitive element in $\F_{q^n}$ with respect to the basis $B$. 
Finally, observe that
$ \langle S \rangle \cup \{0\} $ is isomorphic to the field $\F_{q^n}$
where the addition and multiplication are the standard ones in the ring of square matrices.
In this special case we may already ask whether there are linear $\MRD$ codes
which are not isomorphic to finite fields.

In what follows we do not insist that everything is new. Many facts may have been proved earlier or are at least folklore in the community of
specialists on finite quasifields/semifields/division algebras. However the link between rank metric codes with special parameters and quasifields/semifields/division 
algebras does not seem to have been pointed out in the
existing literature so far.

\section{On the structure of MRD codes with $k=1$ and $n=m$} \label{2}

In this section we connect $\MRD$ codes in case $k=1$ and $m=n$ with well-known objects in finite geometry. Recall that an $\MRD$ code $\mathcal{C}$ in $(\F_q)_{n,n}$ of minimum distance
$n$ is a maximal set of matrices such that $\det (A-B) \not=0$ for all $A \not= B$ in $\mathcal{C}$. Replacing $\mathcal{C}$ by the translate $ \mathcal{C} - B = \{A -B \mid
A \in \mathcal{C}\}$ for some fixed $B \in \mathcal{C}$ we may assume that the zero matrix is an element of $\mathcal{C}$. Hence all matrices in
$\mathcal{C}$ different from zero are invertible.
Replacing $\mathcal{C}$ by $B^{-1}\mathcal{C}$ for some $B \not= 0$ in $\mathcal{C}$ we may further assume that the identity matrix $I$ is an element in $\mathcal{C}$. 
So far we have changed $\mathcal{C}$ by a rank metric distance preserving isometry which is not linear if $0 \not\in \mathcal{C}$.

Furthermore, since $|\mathcal{C}|= q^n$ and $\det A \not=0$ for all $0 \not= A \in \mathcal{C}$ we see that $\mathcal{C}\setminus\{0\}$ acts regularly on the non-zero vectors of
$W=\F_q^n$, i.e. $\mathcal{C}\setminus\{0\}$ acts transitively without fixed points on the non-zero vectors of $W$. 
In particular, if we fix a vector $0 \not= w_0 \in W$, then for any $w \in W$ there
exists exactly one $A(w) \in \mathcal{C}$ such that $ w_0A(w) = w$. In the following we always take $w_0 =e_1 =(1,0,\ldots,0)$. Thus the first row of $A(w)$ is equal to $w$.
In particular, we may write
\begin{equation} \label{e1} \mathcal{C} = \{ A(w) \mid w \in W\}, \end{equation}
where $A(0) =0$ and $A(e_1)=I$. The latter follows by the fact that there is a $w \in W$ with $A(w)=I$ and $\det (A(e_1) - A(w)) = 0$ since the first row of $A(e_1)$ 
and $A(w)=I$ coincide.

In finite geometry, such a system of linear maps is called a spreadset in $W$ (see \cite{Dem,Geometry}), or a spreadset over $\F_q$.
Note that conversely a spreadset in $(\F_q)_{n,n}$ defines an $\MRD$ code $\mathcal{C}$ in $(\F_q)_{n,n}$ with minimum distance
$n$. Spreadsets in $W$ give rise to a multiplication $\circ$ on $W$ defined by 
\begin{equation} \label{e2}  w \circ w' = wA(w') \end{equation}
for $w,w' \in W$.

With this multiplication and the standard vector addition $W$ carries the structure of a quasifield (\cite{Dem}, section 5.1) which is defined as follows.

\begin{definition} \label{def1} {\rm a) A set $\mathcal{Q}$ with two operations $+,\circ:\mathcal{Q} \times \mathcal{Q} \longrightarrow \mathcal{Q}$ is called a (right) quasifield if the following holds.
\begin{itemize}
\item[\rm (i)] $(\mathcal{Q}, +)$ is an abelian group with neutral element $0$ which satisfies $0\circ a =0 = a \circ 0$ for all $a \in \mathcal{Q}$. 
\item[\rm (ii)] There is an identity $e$ in $\mathcal{Q}$ such that $e \circ a = a \circ e = a $ for all $a \in \mathcal{Q}$.
\item[\rm (iii)] For all $a,b \in \mathcal{Q}$ with $a \not=0$ there exists exactly one $x \in \mathcal{Q}$ such that $ a \circ x = b$.
\item[\rm (iv)] For all $a,b,c \in \mathcal{Q}$ with $ a \not=b$ there exists exactly one $x \in \mathcal{Q}$ such that $ x \circ a = x \circ b + c$. 
\item[\rm (v)] $ (a + b) \circ c = a \circ c + b \circ c$ for all $ a,b,c \in \mathcal{Q}$ (right distributivity). 
\end{itemize}
b) We call $$ \kker \mathcal{Q} = \{ c \in \mathcal{Q} \mid c \circ (a +b) = c \circ a + c 
\circ b, \, c \circ (a \circ b) = (c \circ a) \circ b \ \mbox{for all} \ a,b \in 
\mathcal{Q} \}$$ the kernel of the quasifield $\mathcal{Q}$.\\
c) A quasifield $\mathcal{Q}$ which satisfies also the left distributivity law is called a semifield $\mathcal{S}$. 
If $\mathcal{S}$ is not a field, we say that $\mathcal{S}$ is a proper semifield.\\
d) A quasifield with associative multiplication is called a nearfield. In particular, the non-zero elements of a nearfield form a group with respect to $\circ$.\\
}
\end{definition}

\begin{definition} {\rm Let $\mathcal{S}$ be a semifield. \\
a) The left, middle and right nucleus of $\mathcal{S}$ are defined as follows: 
$$\begin{array}{ccc} \mathcal{N}_l = \mathcal{N}_l(\mathcal{S}) = \{ x \in \mathcal{S} \mid x \circ (a \circ b) = (x \circ a) \circ b \ \mbox{for all} \ a,b \in \mathcal{S} \} \\
           \mathcal{N}_m = \mathcal{N}_m(\mathcal{S}) = \{ x \in \mathcal{S} \mid a \circ (x \circ b) = (a \circ x) \circ b \ \mbox{for all} \ a,b \in \mathcal{S} \} \\
		  \mathcal{N}_r = \mathcal{N}_r(\mathcal{S}) = \{ x \in \mathcal{S} \mid a \circ (b \circ x) = (a \circ b) \circ x \ \mbox{for all} \ a,b \in \mathcal{S} \}. 
\end{array}$$
Note that the left nucleus of $\mathcal{S}$ is just the kernel of $\mathcal{S}$ considered as a quasifield.\\
b) The center $Z(\mathcal{S})$ of $\mathcal{S}$ is the set
$$ Z(\mathcal{S}) = \{ a \in \mathcal{N}_l \cap \mathcal{N}_m \cap \mathcal{N}_r \mid x \circ a = a \circ x \ \mbox{for all} \ x \in\mathcal{S} \}.$$
}
\end{definition}

For applications in coding theory we may assume and will do so for the rest 
of this paper that the quasifields, semifields resp. nearfields are always finite. 

\begin{remark} {\rm Quasifields are strongly related to translation planes in finite geometry, i.e., translation planes are precisely those affine planes
which can be coordinatized by quasifields \cite{Dem}. Unfortunately, there is no satisfactory classification of finite quasifields.
In contrast, for finite semifields there is a vast literature \cite{K,Kantor,CW,LP,JJB}. Proper finite semifields exist exactly for all orders 
$p^n \geq 16$ where $p$ is a prime and $ n \geq 3$ \cite{K}. All finite semifields of order $2^5$ have been classified by Walker \cite{W}, of order $3^4$ by Dempwolff \cite{UD}, of order
$2^6$ and $3^5$ by R\'{u}a, Combarro and Ranilla \cite{RCR09},\cite{RCR12}.
Furthermore, finite nearfields have been classified by Zassenhaus in \cite{Z}. 
}
\end{remark}

To continue we recall the following well-known facts which are easy to see.

\begin{lemma} \label{ll1} a) If $\mathcal{Q}$ is a finite quasifield, then $\kker \mathcal{Q}$ is a finite field. \\
b) $\mathcal{Q}$ is a finite dimensional left vector space over $\kker \mathcal{Q}$. \\
c) If $\mathcal{ S}$ is a finite semifield, then $\mathcal{S}$ is a division algebra over its center $Z(\mathcal{S})$. 
\end{lemma}

Now let $\mathcal{Q}$ be a finite quasifield and let
 $ K $ be a subfield of $\kker \mathcal{Q}$. According to Lemma \ref{ll1} we have $\dim_K \mathcal{Q} =n$ for some $n \in \N$.
For $a \in \mathcal{Q}$ we consider the map $ x \mapsto x \circ a$ on $\mathcal{Q}$. Since
$$ (x + y)\circ a = x \circ a + y \circ a \ \mbox{and} \ (k \circ x) \circ a = k \circ (x \circ a) $$
for all $x,y \in \mathcal{Q}$ and all $k \in K$ there exists a unique $R(a)=R_K(a) \in \GL(n,K)$ for $a\not=0$ and $R(0)=0 \in (K)_{n,n}$ such that
$$  xR(a) = x \circ a$$ for all $x \in \mathcal{Q}$.
With this notation the set
$$ \mathcal{C} = \{ R(a) \mid a \in \mathcal{Q} \}$$
is an $\MRD$ code in $(K)_{n,n}$ with minimum distance $n$. Note that $\mathcal{C}$ is uniquely determined by $\mathcal{Q}$ and the chosen subfield $K$ of $\kker \mathcal{Q}$ up to conjugation in $\GL(n,K)$. We will always choose the identity $e \in \mathcal{Q}$ as the first basis vector in a basis of $\mathcal{Q}$ over $K$, hence $e = e_1 = (1,0, \ldots,0)$. Therefore, the first
row in $R(a)$ is equal to $a$ as a vector. 

Conversely, let $\mathcal{C} = \{ A(w) \mid w \in W\}$ be an $\MRD$ code in $(K)_{n,n}$ with minimum distance $n$ as in (\ref{e1}) 
and let $W= \mathcal{W}$ carry the structure of a quasifield defined as in (\ref{e2}). We shall prove that $K \cong K e_1$ is contained in $\kker \mathcal{W}$.
To see that note that
$$ (ke_1)A(w) + (ke_1)A(w') = (ke_1)(A(w) + A(w')) = (ke_1)A(w +w') $$
for $w,w' \in W$ and $k \in K$. Thus 
$$ ke_1 \circ w + ke_1 \circ w' = ke_1 \circ (w+w').$$
Furthermore,
$$  (ke_1 \circ w) \circ w' = (ke_1A(w)) A(w') = k((e_1 A(w))A(w')) = k(w \circ w') 
 = ke_1 \circ (w \circ w'),
 $$
which proves the claim.
Thus we have proved the following which is essentially already stated in \cite{A,BB}.

\begin{theorem} \label{T1} $\MRD$ codes in $(K)_{n,n}$ (containing the zero and identity matrix) with minimum distance $n$ correspond (in the above sense) to finite
quasifields $\mathcal{Q}$ with $K \leq \kker \mathcal{Q}$ and $\dim_{K} \mathcal{Q}= n.$
\end{theorem}

If we require that the codes are closed under addition, hence form abelian groups (since they are finite), we get the following.

\begin{theorem} \label{T2} Additively closed $\MRD$ codes in $(K)_{n,n}$ (containing the identity matrix) with minimum distance $n$ correspond (in the above sense) to finite
semifields $\mathcal{S}$ with $K \leq \kker \mathcal{S}$ and $\dim_{K} \mathcal{S}= n.$
\end{theorem}
\proof{} Suppose that $\mathcal{S}$ is a finite semifield with $K \leq \kker \mathcal{S}$ and $\dim_{K} \mathcal{S}= n$.
Let $\mathcal{C} = \{ R(a) = R_{K}(a) \mid a \in \mathcal{S} \}$.
Since $\mathcal{S}$ satisfies the left distributive law we have 
$$ xR(a+b) = x\circ (a+b) = x\circ a + x\circ b = xR(a) + xR(b) = x(R(a) + R(b)) $$
for $x,a,b \in \mathcal{S}$, hence $R(a)+ R(b) = R(a+b)$.

Conversely, suppose that $\mathcal{Q}$ is a finite quasifield and $\mathcal{C} = \{ R(a) \mid a \in \mathcal{Q} \}$
is additively closed.
Thus, for $a,b \in \mathcal{Q}$ there exists a unique $c \in\mathcal{Q}$ such that
$$ R(a) + R(b) = R(c).$$
If $e$ is the identity in $\mathcal{Q}$ then
\[
	c = e\circ c = eR(c) = e(R(a) + R(b)) = eR(a) + eR(b) = e\circ a + e\circ b = a+b\text{.}
\]
So
\[
	R(a) + R(b) = R(a + b)
\]
or in other words, $R$ is additive.
Thus
$$ x\circ a + x \circ b = xR(a) + xR(b) = x(R(a) +R(b)) = xR(a+b) = x \circ (a+b) $$
for all $x,a,b \in \mathcal{Q}$. This shows that $\mathcal{Q}$ is left distributive, hence $\mathcal{Q}$ is a semifield $\mathcal{S}$.
The fact that $K \leq \kker \mathcal{S}$ and $\dim_{K} \mathcal{S}= n$ follows from Theorem \ref{T1}.
\hfill \qed

In order to understand linearity of $\MRD$ codes over some field $\F_q$ we need the following result.

\begin{proposition} \label{prop1} Let $\mathcal{S}$ be a finite semifield and $K$ a subfield of $\mathcal{S}$.
The following two conditions are equivalent:
\begin{enumerate}
  \item $\mathcal{S}$ is a division algebra over $K$.
  \item $K$ is a subfield of $ Z(\mathcal{S})$.
\end{enumerate}	
\end{proposition}	

\proof{} b) $\Longrightarrow $ a): This follows from Lemma \ref{ll1}. \\
a) $\Longrightarrow $ b): Since $\mathcal{S}$ is a division algebra over $K$, we have
\begin{equation} \label{division algebra} (k \circ a) \circ b = k \circ (a \circ b) = a \circ (k \circ b) \end{equation} 
for all $k\in K$ and all $a,b \in \mathcal{S}$.

The first equality implies $K \leq \mathcal{N}_l(\mathcal{S})$.
Plugging $b = e$ into the equality $(k \circ a) \circ b = a\circ (k\circ b)$ we get
\begin{equation} \label{com}
	k\circ a = a\circ k
\end{equation}
for all $k\in K$ and $a\in\mathcal{S}$.
Hence $K \subseteq \{ x \in \mathcal{S} \mid x \circ a = a \circ x \ \mbox{for all} \ a \in \mathcal{S} \}.$
It remains to show that $ K \subseteq \mathcal{N}_m \cap \mathcal{N}_r$. In order to see that note that
$$ \begin{array}{rcl} a \circ (k \circ b) & = & (k \circ a) \circ b \qquad \qquad \mbox{(by (\ref{division algebra}))} \\
                   & = & (a \circ k) \circ b  \qquad \qquad \mbox{(by (\ref{com}))}
\end{array} $$
for all $k\in K$ and all $a,b \in \mathcal{S}$, hence $K \subseteq \mathcal{N}_m$.
Furthermore,
$$ \begin{array}{rcl}
a \circ (b \circ k) & = & a \circ (k \circ b) \qquad \qquad \mbox{(by (\ref{com}))}	\\
& = & k \circ ( a \circ b) \qquad \qquad \mbox{(by (\ref{division algebra}))}	\\
& = & (a \circ b) \circ k \qquad \qquad \mbox{(by (\ref{com}))}	
\end{array} $$
for all $k\in K$ and all $a,b \in \mathcal{S}$, hence $K \subseteq \mathcal{N}_r$.																
\hfill\qed

\begin{theorem} \label{T3} $K$-linear $\MRD$ codes in $(K)_{n,n}$ (containing the identity matrix) with minimum distance $n$ correspond to finite
division algebras $\mathcal{D}$ over $K$ where $K \leq Z(\mathcal{D})$ and $\dim_{K} \mathcal{D}= n.$
\end{theorem}
\proof{} Let
 $\mathcal{S}$ be a finite semifield with $K \leq \kker \mathcal{S}$ and $\dim_{K} \mathcal{S}= n$. Let $R: \mathcal{S} \rightarrow (K)_{n,n}$ be defined as above; i.e.
$xR(a) = x \circ a$ for $x,a \in \mathcal{S}$ and let $\mathcal{C} =\{R(a) \mid a \in \mathcal{S} \}$ be the $\MRD$ code corresponding to $\mathcal{S}$.
 Clearly, if $R$ is $K$-linear, i.e. $R(k\circ a) = kR(a)$ for $k \in K$ and $a \in \mathcal{S}$, then 
 $\mathcal{C} =\{R(a) \mid a \in \mathcal{S} \}$ is a $K$-vector space. Conversely, if $\mathcal{C}$ is a $K$-vector space, then
 $R(k\circ a) = kR(a)$ for $k \in K$ and $a \in \mathcal{S}$ since the first row of $R(k \circ a)$ and $kR(a)$ coincide.

The condition 
$$  kR(a) = R(k\circ a) $$
for all $k\in K$ and all $a \in \mathcal{S}$ is equivalent to
$$ (kx)R(a) = k(xR(a)) = x(kR(a)) = xR(k\circ a) $$
for all $k\in K$ and all $x,a \in \mathcal{S}$, hence to
$$ (k\circ x) \circ a = k\circ (x \circ a) = x \circ (k \circ a). $$
for all $k\in K$ and all $x,a \in \mathcal{S}$.
The latter means exactly that $\mathcal{S}$ is a division algebra over $K$ and the condition $K \leq Z(\mathcal{S})$ follows by Lemma \ref{ll1}.
\hfill\qed
 
According to the above theorems, division algebras, semifields, nearfields or even quasifields deliver new methods to construct $\MRD$ codes which are different from
Gabidulin codes. Remember that for a Gabidulin code in $(\F_q)_{n,n}$ with minimum distance $n$ the corresponding quasifield is a field.

\section{Isotopy and equivalence}

As stated in Theorem \ref{T3}, $K$-linear $\MRD$ codes in $(K)_{n,n}$ with minimum distance $n$ correspond to finite division algebras $\mathcal{D}$ over $K \leq Z(\mathcal{D})$
with $ \dim_K \mathcal{D} = n$. 
Since non-isomorphic division algebras may lead to equivalent codes we need the following definition.

\begin{definition} {\rm Let $\mathcal{Q}$ and $\mathcal{Q}'$ be finite quasifields which are left vector spaces over the same field $K \leq \kker \mathcal{Q} \cap \kker \mathcal{Q}'$. 
We say that $\mathcal{Q}'$ is isotopic to $\mathcal{Q}$ over $K$ if 
 there are $K$-linear isomorphisms 
$F,G,H: \mathcal{Q} \longrightarrow \mathcal{Q}'$ 
such that
$$  aF \circ' bG = (a \circ b)H $$
for all $a,b \in \mathcal{Q}$. 
}
\end{definition}

Note that the above definition is not standard. Usually isotopy is only defined  over the prime field. Our definition is motivated by Theorem  \ref{equiv} below which states a
very precise and nice connection between isotopy classes of division algebras over $K$ and equivalence classes of their corresponding codes.

\begin{example} \label{ex1} \mbox{} \\
 {\rm
a) There exist exactly $23$ non-isomorphic proper semifields of order $16$ which crumble away into two isotopy classes over $\F_2$ (see \cite{K}, section 6.2). \\
b) Using MAGMA \cite{magma} we computed exactly three equivalence classes of $\MRD$ codes in $(\F_2)_{4,4}$ with minimum distance $4$.
One of these classes represents a Gabidulin code which is associated to the finite field $\F_{16}$. The other two classes are represented as follows (without the zero matrix):\\[2ex]
Code 2:
\begin{center}
{\footnotesize
$ \left(\begin{array}{*{4}{@{\;}c}@{\;}} 1 & 1 & 1 & 0\\ 0 & 1 & 0 & 1 \\ 1 & 1 & 0 & 1 \\ 0 & 1 & 0 & 0\end{array}\right)$,
$\left(\begin{array}{*{4}{@{\;}c}@{\;}} 1 & 0 & 0 & 0\\ 0 & 1 & 0 & 0 \\ 0 & 0 & 1 & 0 \\ 0 & 0 & 0 & 1\end{array}\right)$,
$\left(\begin{array}{*{4}{@{\;}c}@{\;}} 1 & 0 & 1 & 0\\ 1 & 0 & 0 & 1 \\ 1 & 1 & 0 & 0 \\ 0 & 1 & 1 & 1\end{array}\right)$,
$\left(\begin{array}{*{4}{@{\;}c}@{\;}} 0 & 1 & 1 & 0\\ 0 & 0 & 0 & 1 \\ 1 & 1 & 1 & 1 \\ 0 & 1 & 0 & 1\end{array}\right)$,
$\left(\begin{array}{*{4}{@{\;}c}@{\;}} 1 & 1 & 0 & 1\\ 1 & 1 & 1 & 1 \\ 0 & 1 & 1 & 0 \\ 1 & 0 & 0 & 0\end{array}\right)$, \\
$\left(\begin{array}{*{4}{@{\;}c}@{\;}} 1 & 0 & 0 & 1\\ 0 & 0 & 1 & 1 \\ 0 & 1 & 1 & 1 \\ 1 & 0 & 1 & 1\end{array}\right)$,
$\left(\begin{array}{*{4}{@{\;}c}@{\;}} 0 & 0 & 1 & 1\\ 1 & 0 & 1 & 0 \\ 1 & 0 & 1 & 1 \\ 1 & 1 & 0 & 0\end{array}\right)$,
$\left(\begin{array}{*{4}{@{\;}c}@{\;}} 0 & 0 & 1 & 0\\ 1 & 1 & 0 & 1 \\ 1 & 1 & 1 & 0 \\ 0 & 1 & 1 & 0\end{array}\right)$, 
$\left(\begin{array}{*{4}{@{\;}c}@{\;}} 0 & 1 & 0 & 1\\ 1 & 0 & 1 & 1 \\ 0 & 1 & 0 & 0 \\ 1 & 0 & 0 & 1\end{array}\right)$, 
$\left(\begin{array}{*{4}{@{\;}c}@{\;}} 1 & 1 & 1 & 1\\ 0 & 0 & 1 & 0 \\ 1 & 0 & 0 & 0 \\ 1 & 1 & 1 & 0\end{array}\right)$, \\ 
$\left(\begin{array}{*{4}{@{\;}c}@{\;}} 1 & 1 & 0 & 0\\ 1 & 0 & 0 & 0 \\ 0 & 0 & 1 & 1 \\ 0 & 0 & 1 & 0\end{array}\right)$,
$\left(\begin{array}{*{4}{@{\;}c}@{\;}} 0 & 1 & 0 & 0\\ 1 & 1 & 0 & 0 \\ 0 & 0 & 0 & 1 \\ 0 & 0 & 1 & 1\end{array}\right)$,
$\left(\begin{array}{*{4}{@{\;}c}@{\;}} 1 & 0 & 1 & 1\\ 1 & 1 & 1 & 0\\ 1 & 0 & 0 & 1 \\ 1 & 1 & 0 & 1\end{array}\right)$,
$\left(\begin{array}{*{4}{@{\;}c}@{\;}} 0 & 0 & 0 & 1\\ 0 & 1 & 1 & 1 \\ 0 & 1 & 0 & 1 \\ 1 & 0 & 1 & 0\end{array}\right)$,
$\left(\begin{array}{*{4}{@{\;}c}@{\;}} 0 & 1 & 1 & 1\\ 0 & 1 & 1 & 0 \\ 1 & 0 & 1 & 0 \\ 1 & 1 & 1 & 1\end{array}\right).$
}
\end{center}

\noindent
Code 3:
\begin{center}
{\footnotesize
$\left(\begin{array}{*{4}{@{\;}c}@{\;}} 0 & 1 & 1 & 1\\ 1 & 1 & 0 & 0 \\ 0 & 1 & 1 & 0 \\ 0 & 1 & 0 & 0\end{array}\right)$,
$\left(\begin{array}{*{4}{@{\;}c}@{\;}} 1 & 1 & 1 & 0\\ 1 & 1 & 0 & 1 \\ 1 & 1 & 0 & 0 \\ 1 & 0 & 1 & 0\end{array}\right)$,
$\left(\begin{array}{*{4}{@{\;}c}@{\;}} 1 & 0 & 0 & 0\\ 0 & 1 & 0 & 0 \\ 0 & 0 & 1 & 0 \\ 0 & 0 & 0 & 1\end{array}\right)$,
$\left(\begin{array}{*{4}{@{\;}c}@{\;}} 1 & 0 & 0 & 1\\ 0 & 0 & 0 & 1 \\ 1 & 0 & 1 & 0 \\ 1 & 1 & 1 & 0\end{array}\right)$, 
$\left(\begin{array}{*{4}{@{\;}c}@{\;}} 0 & 1 & 0 & 0\\ 0 & 0 & 1 & 0 \\ 0 & 0 & 0 & 1 \\ 1 & 1 & 0 & 0\end{array}\right)$, \\
$\left(\begin{array}{*{4}{@{\;}c}@{\;}} 1 & 1 & 0 & 1\\ 0 & 0 & 1 & 1 \\ 1 & 0 & 1 & 1 \\ 0 & 0 & 1 & 0\end{array}\right)$,
$\left(\begin{array}{*{4}{@{\;}c}@{\;}} 0 & 1 & 1 & 0\\ 1 & 0 & 0 & 1 \\ 1 & 1 & 1 & 0 \\ 1 & 0 & 1 & 1\end{array}\right)$,
$\left(\begin{array}{*{4}{@{\;}c}@{\;}} 0 & 0 & 1 & 0\\ 1 & 0 & 1 & 1 \\ 1 & 1 & 1 & 1 \\ 0 & 1 & 1 & 1\end{array}\right)$,
$\left(\begin{array}{*{4}{@{\;}c}@{\;}} 1 & 0 & 1 & 0\\ 1 & 1 & 1 & 1 \\ 1 & 1 & 0 & 1 \\ 0 & 1 & 1 & 0\end{array}\right)$, 
$\left(\begin{array}{*{4}{@{\;}c}@{\;}} 1 & 1 & 0 & 0\\ 0 & 1 & 1 & 0 \\ 0 & 0 & 1 & 1 \\ 1 & 1 & 0 & 1\end{array}\right)$, \\
$\left(\begin{array}{*{4}{@{\;}c}@{\;}} 1 & 0 & 1 & 1\\ 1 & 0 & 1 & 0 \\ 0 & 1 & 0 & 1 \\ 1 & 0 & 0 & 1\end{array}\right)$,
$\left(\begin{array}{*{4}{@{\;}c}@{\;}} 0 & 0 & 0 & 1\\ 0 & 1 & 0 & 1 \\ 1 & 0 & 0 & 0 \\ 1 & 1 & 1 & 1\end{array}\right)$,
$\left(\begin{array}{*{4}{@{\;}c}@{\;}} 0 & 1 & 0 & 1\\ 0 & 1 & 1 & 1 \\ 1 & 0 & 0 & 1 \\ 0 & 0 & 1 & 1\end{array}\right)$, 
$\left(\begin{array}{*{4}{@{\;}c}@{\;}} 1 & 1 & 1 & 1\\ 1 & 0 & 0 & 0 \\ 0 & 1 & 0 & 0 \\ 0 & 1 & 0 & 1\end{array}\right)$,
$\left(\begin{array}{*{4}{@{\;}c}@{\;}} 0 & 0 & 1 & 0\\ 1 & 1 & 1 & 0 \\ 0 & 1 & 1 & 1 \\ 1 & 0 & 0 & 0\end{array}\right)$.
}
\end{center}

The Codes 2 and 3 correspond naturally to the two non-isotopic semifields of order 16 as the next result shows. Thus these matrices can be obtained easily from the known semifields.

}
\end{example}

Over the prime field the next result is stated explicitly in (\cite{MP}, Proposition 2.1). The reader may also compare Section 1.2 and 1.3 of \cite{LP} with
the following statement.

\begin{theorem} \label{equiv} Let $\mathcal{C}, \mathcal{C}' \subseteq (K)_{n,n}$ be $K$-linear $\MRD$ codes with minimum distance $n$ and corresponding division algebras ${\mathcal{D}}$ and 
${\mathcal{D}}'$. In particular $K \leq Z(\mathcal{D})$ and $K \leq Z(\mathcal{D}')$.
Then $\mathcal{C}$ and $\mathcal{C}'$ are linearly equivalent if and only if $\mathcal{D}'$ is isotopic to $\mathcal{D}$ or its transpose $\mathcal{D}^t$ over $K$.
\end{theorem}
\proof{} If $W= K^n$ then $\mathcal{C} = \{ A(w) \mid w \in W\}$ where $e_1A(w) = w$ for $w \in W$. 
Furthermore, if ${\mathcal{D}}$ is the corresponding division algebra where ${\mathcal{D}} = W$ as a $K$-vector space, then the multiplication
on ${\mathcal{D}}$ is given by
$$   w_1 \circ w_2 = w_1 A(w_2) \quad \mbox{for} \ w_i \in W.$$
The transpose $\mathcal{D}^{t}$ of $\mathcal{D}$ is defined by $\mathcal{D}^{t} =W$ as a $K$-vector space but with multiplication $w_1 \circ w_2 = w_1A(w_2)^t$. 
We use the same notation for the second code but with a $'$ everywhere. \\[1ex]
We first suppose that $\mathcal{C}$ and $\mathcal{C}'$ are equivalent and prove that the corresponding semifields are isotopic over $K$.
Thus by assumption there exist $X, Y \in \GL(n,K)$
such that
$$  \{ XA(w)Y \mid w \in W \}= \{ A'(w) \mid w \in W\} \qquad \qquad (*) $$
or
$$ \{ XA(w)Y \mid w \in W \}= \{ A'(w)^t \mid w \in W\}. \qquad \qquad (**)$$
Suppose that $(*)$ holds true.
This means that for each $w \in W$ there exists exactly one $\tilde{w} \in W$ such that
$$  XA(w)Y = A'(\tilde{w}).$$
Let $F: W \longrightarrow W$ denote the map $wF= \tilde{w}$. For $w_1,w_2 \in W$ we obtain
$$ (w_1 \circ w_2)Y = w_1A(w_2)Y = w_1X^{-1}A'(\tilde{w_2}) = w_1 X^{-1}A'(w_2F) = w_1X^{-1} \circ' w_2F.$$
Since $Y^{-1}$ and $ X$ are $K$-linear it remains to show that $F$ is $K$-linear as well. 

First note that $A(k_1w_1 + k_2w_2) = k_1A(w_1) + k_2A(w_2)$ for $k_i \in \F_q$ and $w_i \in W$ since the first row of $A(w)$ is equal to $w$ and $\mathcal{C}$ is 
a $K$-vector space.
The same holds for $A'$.
From this we obtain
$$ \begin{array}{rcl} A'((k_1w_1 + k_2w_2)F) & = & XA(k_1w_1 + k_2w_2)Y \\[1ex]
	& = & k_1XA(w_1)Y +k_2XA(w_2)Y \\[1ex]
	& = & k_1A'(w_1F) + k_2A'(w_2F) \\[1ex]
	& = & A'( k_1(w_1F) + k_2(w_2F)).
   \end{array}
$$																			
Applying the inverse of $A'$ we get
$$ (k_1w_1 + k_2w_2)F = k_1(w_1F) + k_2(w_2F) $$
which proves that $F$ is $K$-linear. Thus the division algebras ${\mathcal{D}}$ and ${\mathcal{D}}'$ are isotopic over $K$. 
In case $(**)$ the proof runs similar.\\

Now suppose that the corresponding division algebras are isotopic over $K$.	Thus there are $K$-linear isomorphisms $F,G,H: W \longrightarrow W$ such that
$$ (w_1 \circ w_2)H = w_1 F\circ' w_2 G $$	
for $v,w \in W$. If follows
$$ w_1A(w_2)H = w_1FA'(w_2G)$$
for all $w_i \in W$. This implies $A(w)H = FA'(wG)$ for $w \in W$ or 
$$ F^{-1}A(w)H = A'(wG)$$ 
for all $w \in W$. Thus
$$ \begin{array}{rcl} F^{-1}\mathcal{C}H & = & \{F^{-1}A(w)H \mid w \in W\}\\[1ex]
       & = & \{A'(wG) \mid w \in W\} \\[1ex]
						 & = & \{ A'(w) \mid w \in W\} = \mathcal{C}'.
	\end{array}					
$$
This shows that $\mathcal{C}$ and $\mathcal{C}'$ are equivalent.
The case that $\mathcal{D}$ is isotopic to the transpose of $\mathcal{D'}$ is done similarly.
\hfill\qed

\begin{remark} {\rm  
a) The second part of the proof of Theorem \ref{equiv} shows that an isotopy between two quasifields always leads to equivalent codes. \\
b)  One of the anonymous referees of this paper informed us that Proposition 2.1 and Theorem 2.2 of \cite{MP} prove the following.\\
Let $\mathcal S$ and ${\mathcal S'}$ be finite semifields with corresponding (additively closed) $\MRD$ codes ${\mathcal C, C'} \subseteq (K)_{n,n}$. Then $\mathcal C$ and $\mathcal C'$
are equivalent (in the sense of Theorem \cite{Hua} including automorphisms of $K$) if and only if $\mathcal S$ and $\mathcal S'$ are isotopic over the prime field of $K$.
}
\end{remark}

\section{Symmetric MRD codes}

Let $E=\F_{q^n}$ and let $K =\F_q \leq E$. On $E$ the standard non-degenerate symmetric $K$-bilinear form $ \langle \cdot \, ,\cdot \rangle$ is defined by
$$  \langle x,y \rangle \, = \, \tr_{E/K}(xy) $$
for $x,y \in E$ where $\tr$ denotes the trace of $E$ over $K$. If $a$ is running over all non-trivial elements of $E$ we get non-degenerate symmetric $K$-bilinear forms
of $E$ by
$$ \langle x,y \rangle_a \, = \, \langle a x,y \rangle.$$
Taking the corresponding Gram matrices together with the zero matrix we obtain a linear $\MRD$ code in $(\F_q)_{n,n}$ with minimum distance $n$ consisting of symmetric matrices.
This code is equivalent to a Gabidulin code which can be seen as follows. We fix a basis $x_1, \ldots, x_n$ of $E$ over $K$. Let $y_i = Bx_i $ 
for $i=1, \ldots,n$ be the dual basis. Then the Gram matrices with respect to the basis $x_1, \ldots, x_n$ are of the form $A^tB^{-1}$ where $A$ runs through
a Singer subgroup of $\GL(n,q)$.

According to \cite{GP}, the symmetry can be used to correct (symmetric) errors beyond the bound $\lfloor \frac{d-1}{2} \rfloor$.

\begin{definition} {\rm Let $K$ be a finite field. We call a code $\mathcal{C} \subseteq (K)_{n,n}$ symmetric if all matrices $A$ in $\mathcal{C}$ are symmetric, i.e., $A=A^t$
for all $ A \in \mathcal{C}$.
}
\end{definition}

\begin{definition} {\rm Let $\mathcal{Q}$ be a finite quasifield over $K \leq \kernel \mathcal{Q}$. A $K$-bilinear form
$ \langle \cdot \, ,\cdot \rangle$ on $\mathcal{Q}$ is called invariant if
$$  \langle x \circ a, y \rangle \, = \, \langle x, y\circ a \rangle $$
for all $a,x,y \in \mathcal{Q}$.
}
\end{definition}

\begin{lemma} \label{sy} Let $\mathcal{Q}$ be a finite quasifield over $K \leq \kernel \mathcal{Q}$ and let $ \langle \cdot \, ,\cdot \rangle$ be an invariant non-degenerate
symmetric $K$-bilinear form on $\mathcal{Q}$.  For $ a \in \mathcal{Q}$ we define the form $ \langle \cdot \, ,\cdot \rangle_a$ by
$$ \langle x \, ,y \rangle_a = \langle x \circ a \, , y \rangle $$
for $x,y \in \mathcal{Q}$. 
Then, for all \,$ 0 \not= a \in \mathcal{Q}$, the form  $ \langle \cdot \, ,\cdot \rangle_a$ is $K$-bilinear, non-degenerate and symmetric.
\end{lemma}
\proof{} One easily checks that $ \langle \cdot \, ,\cdot \rangle_a$ is $K$-bilinear since $(x_1 + x_2) \circ a = x_1 \circ a + x_2 \circ a$ and
$ (k \circ x) \circ a = k \circ (x \circ a)$ for all $x, x_1, x_2, a \in \mathcal{Q}$ and all $k \in K$. 

Let $G$ denote the Gram matrix of $ \langle \cdot \, ,\cdot \rangle$ with respect to the basis as $R(a)$ is taken where $x \circ a = xR(a)$.
Then $ \langle \cdot \, ,\cdot \rangle_a $ has the Gram matrix $R(a)^tG$ which is regular for $a \not= 0$. Thus $ \langle \cdot \, ,\cdot \rangle_a$
is non-degenerate for all $0 \not= a \in \mathcal{Q}$.

Finally, the symmetry follows by
$$ \begin{array}{rcl} \langle x \, ,y \rangle_a & = & \langle x \circ a \, , y \rangle \\[1ex]
            & = & \langle x \, ,y \circ a\rangle  \qquad \qquad (\mbox{since} \ \langle \cdot \, ,\cdot \rangle \ \mbox{is invariant})\\[1ex]
											 & = & \langle y \circ a \, ,x \rangle  \qquad \qquad (\mbox{since} \ \langle \cdot \, ,\cdot \rangle \ \mbox{is symmetric}) \\[1ex]
											 & = & \langle y \, ,x \rangle_a  
\end{array} $$
for all $x,y,a\in \mathcal{Q}$.
\hfill\qed

\begin{theorem} \label{symmetric}
Let $\mathcal{Q}$ be a finite quasifield over the field $K \leq \kernel \mathcal{Q}$ and let $\mathcal{C} =\{R(a) \mid a \in \mathcal{Q} \}$ be the corresponding
 $\MRD$ code in $(K)_{n,n}$. Then 
$\mathcal{Q}$ admits an invariant non-degenerate symmetric $K$-bilinear form if and only if the equivalence class of $\mathcal{C}$ contains a symmetric code. 
\end{theorem}
\proof{} Suppose that $\mathcal{Q}$ admits an invariant non-degenerate symmetric $K$-bilinear form $ \langle \cdot \, ,\cdot \rangle$. 
According to Lemma \ref{sy} the $K$-bilinear forms $ \langle \cdot \, ,\cdot \rangle_a$ are non-degenerate and symmetric for $a \not= 0$ with Gram matrices $R(a)^tG$.
Furthermore for $ a \not= b$ in $\mathcal{Q}$, the difference of the corresponding Gram matrices 
$$ R(a)^tG -R(b)^tG = (R(a) -R(b))^tG$$
is regular.
 Thus $\{ R(a)^tG \mid a \in\mathcal{Q}\}$ is a symmetric MRD code which is equivalent to $\mathcal{C}$.
 
Conversely suppose that the equivalence class of $\mathcal{C}$ contains a symmetric code. Thus we may assume that there are regular matrices
 $X$ and $Y$ such that $\{XR(a)Y \mid a \in {\mathcal Q} \}$ consists of symmetric matrices. 
(The second type of equivalence in Theorem \ref{Hua} leads to the same just by taking transpose matrices.)
Since $R(e)$ is the identity matrix we have $(XY)^t =XY$, hence $Z=YX^{-t} =X^{-1}Y^t =Z^t$. 
Let $\langle \cdot \, , \cdot \rangle$ be the standard symmetric non-degenerate bilinear form on the $K$-vector space ${\mathcal Q}$.
Thus the non-degenerate bilinear form $\langle \cdot \, , \cdot \rangle_Z$
defined by
$$ \langle x,y\rangle_Z \, = \, \langle x,yZ \rangle$$ is symmetric. From $(XR(a)Y)^t = Y^t R(a)^tX^t = XR(a)Y$ we get
$$ X^{-1}Y^tR(a)^t =R(a)YX^{-t},$$ hence $ZR(a)^t = R(a)Z$. It follows
$$ \langle x \circ a,y \rangle_Z = \langle xR(a),yZ \rangle = \langle x,yZR(a)^t \rangle = \langle x,yR(a)Z \rangle = \langle x,y \circ a \rangle_Z$$
and $\langle \cdot \, , \cdot \rangle_Z$ is invariant.
\hfill\qed

\begin{remark} {\rm a) In finite geometry Theorem \ref{symmetric} can be stated as follows \cite{Knarr}: The translation plane associated to a quasifield
$\mathcal{Q}$ is symplectic if and only if $\mathcal{Q}$ admits a non-degenerate invariant symmetric bilinear form. \\
b) According to (\cite{Knarr}, Theorem 4.2), $\mathcal{Q}$ admits a non-degenerate invariant symmetric bilinear form over $\kker \mathcal{Q}$ if and only if the
additive group generated by $\{ (xy)z-x(zy) \mid x,y,z \in \mathcal{Q} \}$ is a proper subgroup of $(\mathcal{Q}, +)$. \\
c) Symmetric $\MRD$ codes have been explicitly constructed by Kai-Uwe Schmidt in \cite{Kai}.
}
\end{remark}

\section{Finite nearfields}

Finite nearfields have been classified by Zassenhaus (see \cite{Z}, or section 5.5.2 in \cite{Dem}). These are the regular quasifields, usually denoted by $N(n,q)$, and seven
exceptional cases. The quasifields $N(n,q)$ exist for all $q$ and $n$, provided all prime divisors of $n$ divide $q-1$ and $ 4 \mathop{\not|} n$ in case $ q \equiv 3 \pmod 4$.
Note that $N(n,q)$ is not a field if $n>1$, and $N(1,q) = \F_q$. Moreover, the center of $N(n,q)$ is $\F_q$.
If $\mathcal{C}$ is the $\MRD$ code corresponding to $N(n,q)$ for $n>1$ then $\mathcal{C}^* = \mathcal{C} \setminus \{0\}$ is a nonabelian group of order $q^n-1$ whereas
in the class of Gabidulin codes this group is cyclic of order $q^n-1$.
We demonstrate one of the exceptional cases in the next example.

\begin{example} {\rm Let $Q$ be the subgroup of $GL(2,11)$ generated by
 $ A=\left(\begin{array}{cr} 0 & -1\\ 1 & 0 \end{array} \right)$ and $ B=\left(\begin{array}{cr} 2 & 4\\ 1 & -3 \end{array} \right)$. One easily checks that ${Q} \cong \SL(2,5)$. Furthermore ${Q}$ acts regularly on the non-zero vectors of $V(2,11)$. 
Thus $\mathcal{C} = Q \cup \{0\}$ is an $\MRD$ code in $(\F_{11})_{2,2}$.
The orders of elements of ${Q}$ are $1,2,3,4,5,6$ and $10$. Since $E+A$ has order $40$, the rank metric code $\mathcal{C}$ is not additively closed.
}
\end{example}

\section{MRD codes with $k>1$ and $n=m$}

\begin{example} {\rm Using Magma we see that $(\F_3)_{3,3}$ contains two equivalence classes of linear $\MRD$ codes with minimum distance $d=n-k+1 =2$.
One of them represents the Gabidulin code $\mathcal{G}$ and has the following matrices as a basis.

$$ \left(\begin{array}{*{3}{@{\;}c}@{\;}} 1 & 0 & 0 \\ 0 & 0 & 0 \\ 0 & 1 & 0 \end{array}\right), 
 \left(\begin{array}{*{3}{@{\;}c}@{\;}} 0 & 1 & 0 \\ 0 & 0 & 0 \\ 1 & 2 & 1 \end{array}\right),
 \left(\begin{array}{*{3}{@{\;}c}@{\;}} 0 & 0 & 1 \\ 0 & 0 & 0 \\ 0 & 1 & 2 \end{array}\right),
 \left(\begin{array}{*{3}{@{\;}c}@{\;}} 0 & 0 & 0 \\ 1 & 0 & 0 \\ 0 & 0 & 2 \end{array}\right),
 \left(\begin{array}{*{3}{@{\;}c}@{\;}} 0 & 0 & 0 \\ 0 & 1 & 0 \\ 0 & 2 & 1 \end{array}\right),
 \left(\begin{array}{*{3}{@{\;}c}@{\;}} 0 & 0 & 0 \\ 0 & 0 & 1 \\ 2 & 1 & 0 \end{array}\right). 
$$
\mbox{}\\
The other class contains a code $\mathcal{C}$ which has the following basis. 

$$ \left(\begin{array}{*{3}{@{\;}c}@{\;}} 1 & 0 & 0 \\ 0 & 0 & 0 \\ 1 & 1 & 0 \end{array}\right), 
 \left(\begin{array}{*{3}{@{\;}c}@{\;}} 0 & 1 & 0 \\ 0 & 0 & 0 \\ 0 & 0 & 2 \end{array}\right),
 \left(\begin{array}{*{3}{@{\;}c}@{\;}} 0 & 0 & 1 \\ 0 & 0 & 0 \\ 2 & 0 & 2 \end{array}\right),
 \left(\begin{array}{*{3}{@{\;}c}@{\;}} 0 & 0 & 0 \\ 1 & 0 & 0 \\ 1 & 2 & 1 \end{array}\right),
 \left(\begin{array}{*{3}{@{\;}c}@{\;}} 0 & 0 & 0 \\ 0 & 1 & 0 \\ 2 & 2 & 1 \end{array}\right),
 \left(\begin{array}{*{3}{@{\;}c}@{\;}} 0 & 0 & 0 \\ 0 & 0 & 1 \\ 1 & 2 & 2 \end{array}\right). 
$$	
The only semifield of order $27$ has the Frobenius map $x \mapsto x^3$ in $\F_{27}$ as a semifield automorphism and we may
use the Gabidulin construction over the semifield to get a rank metric code $\mathcal{C}'$. However, up to equivalence we do not get the above code 
since $\mathcal{C}'$ contains matrices of rank $1$. 
Note that the rank distribution of $\mathcal{C}$ and the corresponding Gabidulin code coincide, consistently with Theorem 5.6 of \cite{D}. There are exactly $338$ matrices of rank $2$ and $390$
 of rank $3$. Furthermore, if we consider the matrices of $\mathcal{C}$ as vectors in $\F_{27}^3$ we obtain an MDS code which is not linear over
$\F_{27}$ in contrast to the corresponding Gabidulin code.

According to Delsarte \cite{D} the duals of the codes $\mathcal{G}$ and $\mathcal{C}$ are again $\MRD$ codes, where the duality is defined by
$$ \langle A,B\rangle =\tr (AB^t) $$
for $A,B \in (\F_3)_{3,3}$.
We would like to mention here that both $\mathcal{G}$ and $\mathcal{C}$ are the duals of the codes arising from the two semifields of order $27$.
}
\end{example}

\section*{Acknowledgement}
We would like to thank Peter M\"uller who brought the connection of MRD codes in $(K)_{n,n}$ of minimum distance $n$ with spreadsets to our attention. 
Furthermore, we would like to thank Thomas Honold who told us about the book \cite{Wan} and Hua's result. Finally, the authors appreciated valuable discussions
with John Sheekey at the ALCOMA15 meeting at Banz. Special thanks goes to the anonymous referees which we owe some improvements.

\end{document}